\documentclass{amsart}

\usepackage{enumitem}
\usepackage[T1]{fontenc}
\usepackage[utf8]{inputenc}
\usepackage[english]{babel}
\usepackage{amsmath}
\usepackage{accents}
\usepackage{amsthm}
\usepackage{amssymb}
\usepackage[a4paper, margin=1in]{geometry}
\usepackage{comment}

\usepackage{subcaption}

\usepackage{graphicx}
\usepackage{tikz}
\usetikzlibrary{shapes.misc}
\usepackage{float}

\usepackage[unicode=true,pdfusetitle,
bookmarks=true,bookmarksnumbered=false,bookmarksopen=false,
breaklinks=false,pdfborder={0 0 1},backref=false,hidelinks]
{hyperref}

\makeatletter

\numberwithin{equation}{section}
\numberwithin{figure}{section}

\makeatother

\newlength{\dhatheight}
\newcommand{\doublehat}[1]{%
    \settoheight{\dhatheight}{\ensuremath{\hat{#1}}}%
    \addtolength{\dhatheight}{-0.35ex}%
    \hat{\vphantom{\rule{1pt}{\dhatheight}}%
    \smash{\hat{#1}}}}

\begin{document}
	
	\title[Searching point patterns for local topography]{Searching point patterns in point clouds describing local topography}
	
	\author{E. Bednarczuk$^{1}$ 	}
	\author{R. Bieńkowski$^{1,2}$ 	}	
	\author{R. Kłopotek$^{3}$ 	}
    \author{J. Kryński$^{4}$ 	}
    \author{K. Leśniewski$^{1}$ }
    \author{K. Rutkowski$^{1}$ }
    \author{M. Szelachowska$^{4}$}

    \thanks{$^1$ \textit{Systems Research Institute Polish Academy of Sciences}, Warsaw, Poland	}
    \thanks{$^2$\textit{WIT Academy}, Warsaw, Poland}
    \thanks{$^3$ \textit{Institute of Computer Science, Cardinal Stefan Wyszynski University in Warsaw}, Warsaw, Poland}
	\thanks{$^4$ \textit{Institute of Geodesy and Cartography}, Warsaw, Poland}

	\begin{abstract}
We address the problem of comparing and aligning spatial point configurations in $\mathbb{R}^3$ arising from structured geometric patterns. Each pattern is decomposed into arms along which we define a normalized finite-difference operator measuring local variations of the height component with respect to the planar geometry of the pattern.
This quantity provides a parametrization-independent local descriptor that complements global similarity measures. In particular, it integrates naturally with Wasserstein-type distances for comparing point distributions and with Procrustes analysis for rigid alignment of geometric structures. 

	\end{abstract}
    	\maketitle
	\section{Introduction}

The Earth's features are frequently acquired from the measurements carried out at discrete points. Coordinates of those points must be determined to make the measurements useful. Before the era of GPS technology, the horizontal coordinates of such points were usually obtained from maps, which resulted in their low accuracy. The quantities describing Earth features, mainly of physical character, were measured at those points with quite high accuracy even for the present time. The improvement of the horizontal coordinates of such points by using additional available information about Earth's surface and computational algorithms leads to obtaining data sets much more valuable for geosciences, without the necessity of repeating very long-lasting and expensive measurement campaigns.

In 1957-1979 the whole territory of Poland was roughly homogeneously covered with gravimetric stations at which gravity was measured. The set of over one million points with gravity determined at the accuracy level below $0.1$ {mGal ($10^{-6} m/s^2$) is up to now extremely valuable and unique and is widely used in Earth sciences, especially in geophysics, geodesy, and geology. Heights of gravimetric stations were determined by spirit leveling at the accuracy level of about 5 cm. The locations of gravity stations, i.e. their horizontal coordinates, were determined from printed maps. Errors in these coordinates have several sources, including:
\begin{itemize}
    \item inaccurate identification of a point location,
    \item inaccurate marking of a survey point on a county map in the field,
    \item  inaccurate marking of a point on a topographic map at the scale of 1:50 000,
    \item error in reading coordinates of a point from a topographic map.
\end{itemize}

The inaccurate and sometimes even erroneous location of a gravimetric point is a weakness of the considered gravimetric data set for Poland. It can be expected that the errors of horizontal coordinates of gravimetric stations are over a hundred meters \cite{krynski2007precyzyjne}. In addition, as part of the aforementioned geophysical work, measurements of the course of the local Earth’s surface were carried out at the gravimetric points surrounded by rougher topography. The topographic data around such gravimetric points are given in the form of so-called ’crosses’, i.e., heights along the mutually perpendicular profiles measured up to 100 meters from their intersection at the gravimetric point in each direction. The direction coinciding with the greatest slope was marked as the first direction.

In recent few decades, the terrain heights data has become available in the form of digital elevation models (DEM). The DEM is a discrete representation of the terrain surface in a grid form. In the last decade, a spatial resolution of DEMs has reached 1 m (see \cite{GeoportalDEM2026}) or even below a meter (see \cite{DataGovUK_LIDAR_DTM_2017_25cm}).

With the increasing availability of precise high resolution of DEMs and precisely determined heights of gravimetric points with the information on topography in their vicinity, much more reliable identification of localization of gravimetric points, i.e., their horizontal coordinates, became possible. It can be done by matching 3D surfaces represented by DEMs. In the following research the problem has been approached via Wasserstein metrics and a linear optimization applied to real geospatial data.

For numerical experiments conducted in this research, the unique documentation containing not only the coordinates of the gravity points but also a description of the topography in the vicinity of these points was used (Section \ref{subsec:local_topography}). Details of the DEM utilized are given in Section \ref{subsec:local_topography}. Problem formulation and methods applied in this research are presented in Section \ref{sec:problem}. The algorithm developed is given in Section \ref{sec:algorithm}. The numerical examples are described in Section \ref{sec:computional}.

\section{Data used}\label{sec:data_used}

Point patterns from local topography data were matched with the respective ones from high-resolution DEM. Both the horizontal and vertical coordinates of the points used in the investigation, as well as lengths along profiles, are given in meters. Also, heights used in numerical calculations are given in meters.

\subsection{ Local topography data}\label{subsec:local_topography}
The information about the very local topography around gravimetric points, i.e., within a 100 m radius, is given in a discrete form for each gravimetric point for which the slopes of the terrain in its immediate vicinity exceed 6 degrees \cite{krynski2007precyzyjne}. Depending on the complexity of the topography, the measurements were conducted along the straight lines in four or eight directions evenly spaced in the horizon \cite{krynski2007precyzyjne}. The slopes were measured with the inclinometer, while the distance was measured with the coiled steel tape. The original gravimetric points locations were given as geographical coordinates in the Borowa Góra reference frame and transformed into planar coordinates in PL-1992 reference frame. The height values were given in the Kronsztadt 60 system and transformed to PL-EVRF2007-NH reference frame.  

The 10 cm accuracy of heights and 2 m accuracy of positions of the profile points with respect to the gravimetric point has roughly been estimated. An example of a description of such measurements is shown in Figure 1.

along four mutually perpendicular lines (arms, intervals). This structure will further be referred to as the initialization target.} 

\subsection{High resolution DEM}\label{subsec:high_resolution_DEM}
The digital elevation model (DEM) with the resolution of 1 m (given in 1 m × 1 m grid) was used in this research. The DEM is developed mainly on  the basis of data from airborne laser scanning (ALS). The model can be downloaded from the official Polish government portal \cite{GUGiK2026}. Numerical experiments were performed using 52 sections of the model, each representing 2.3 km × 2.3 km area (Fig. 2). For the area investigated, the accuracy of the DEM heights given by the  data provider, for most of the sections is assessed as 15 cm, with the exception of one section for which the accuracy is at the level of 50 cm and three sections with the accuracy of 20 cm. Heights of the DEM are referred to PL-EVRF2007-NH reference frame, and horizontal coordinates to PL-1992 reference frame.

\section{Problem formulation}\label{sec:problem}

In the process of enhancing the accuracy of point coordinates on a map, the "linear sum assignment" model proves to be a valuable tool. This article explores the application of this model to correct the positions of points whose coordinates are affected by substantial errors. A significant challenge in this task is the limited information available about the terrain surrounding each point given along the slopes in four mutually perpendicular directions. Despite this limitation, the method can still be framed as a matching problem, aiming to align the erroneous coordinates with the most accurate possible positions based on the available, albeit partial, data. This approach, using a more precise representation of the terrain, provides the corrected coordinates that, however, are still biased due to the limited initial information and the known topographic features.

The  patterns \( Q_\ell \) of points are constructed from a point cloud \( \mathcal{P} \) in three-dimensional space with the number of elements $N$. Each point in \( \mathcal{P} \) has coordinates \( (x, y, z) \), and the goal is to select subsets of points from \( \mathcal{P} \) that form structures similar to the given reference  pattern \( T \). The similarity is ensured by suitably chosen criteria. The construction process ensures that \( Q_\ell \) preserves the geometric properties of \( T \).

\subsection{Properties of the reference  pattern \( T \) of points}
The   pattern \( T \) of points from the reference set is defined as:
$$
T := \{ P_i = (x_i, y_i, z_i) \in \mathbb{R}^3 \mid i = 0, \dots, m \},
$$
where  \( t_0 = (x_0, y_0, z_0) \) is referred as the center of the pattern $T$.

The reference point pattern $T_\theta$ of the points of $T$ rotated with respect to z-axes by the angle $\theta$ with coordinates $(x,y)$ referred to the center point $t_0$
\begin{equation*}
    T_\theta := \{ P_i^\theta = (x_i^\theta, y_i^\theta, z_i) \mid i = 0, \dots, m \},
\end{equation*}
where
\begin{align*}
    &[x_i^\theta,y_i^\theta]^T = [x_0,y_0]^T + \operatorname{rot}(\theta) \cdot [x_i-x_0,y_i- y_0]^T,\quad i = 0, \dots, m,\\
    &\operatorname{rot}(\theta) = \left[\begin{array}{cc}
        \cos (\theta) & - \sin (\theta) \\
        \sin (\theta) & \cos (\theta) 
    \end{array}\right].
\end{align*}

has also been introduced.

\subsection{Properties of the sought pattern $Q_\ell$}
The sought for $Q_\ell$, $\ell =\{1,\dots, \binom{N}{m+1}\}$ is a pattern of $\mathbb{R}^3$ points
$$
Q_\ell := \{ q_{\ell,i} = (\tilde{x}_{\ell,i}, \tilde{y}_{\ell,i}, \tilde{z}_{\ell,i}) \in \mathcal{ P} \mid i = 0, \dots, m \},
$$
contained in the cloud $\mathcal{ P}$. The point $q_{\ell,0}=(\tilde{x}_{\ell,0},\tilde{y}_{\ell,0},\tilde{z}_{\ell,0})$ is called the center of  $Q_\ell$.

{\bf Requirements:}
\begin{enumerate}
    \item 
Searched is the $q_{\ell,0}$ satisfying
\begin{equation}\label{ineq:req_z}
 |z_0-\tilde{z}_{\ell,0}|\leq  d_1\ \quad \wedge \quad \| (x_{0}, y_{0} ) - (\tilde{x}_{\ell,0}, \tilde{y}_{\ell,0}) \|_{\infty} \leq r\ , 
\end{equation}
    where \( \| \cdot \|_\infty \) is the $\ell_\infty$ norm, and $r>d_2$ is given.
 
This allows to limit the search for the center of $Q_{\ell}$, $q_{\ell,0}=(\tilde{x}_{\ell,0},\tilde{y}_{\ell,0},\tilde{z}_{\ell,0})$ such that:
\begin{itemize}
    \item  $z_0-d_1\leq \tilde{z}_{\ell,0} \leq z_0+d_1$,
    \item $ x_0-r \leq \tilde{x}_{\ell,0} \leq  x_0+r $ and  $ y_0-r \leq \tilde{y}_{\ell,0} \leq  y_0+r $.
\end{itemize}
\item Moreover, for each $q_{\ell,0}$ satisfying  \eqref{ineq:req_z} a pattern $Q_\ell$  satisfying:
\begin{align}\label{ineq:req_y_theta}
\begin{aligned} 
    \exists_{\theta \in [0,2\pi]} \quad \forall_{i\in \{0,1,\dots,m\} } \quad &[\tilde{x}_{\ell,i},\tilde{y}_{\ell,i}]-[\tilde{x}_{\ell,0},\tilde{y}_{\ell,0}] \approx [x_{i}^\theta,y_{i}^\theta]-[x_{0}^\theta,y_{0}^\theta]=[x_{i}^\theta,y_{i}^\theta]-[x_{0},y_{0}],\\
    & \tilde{z}_{\ell,i}=\operatorname{val}_{\mathcal{ P},z} (\tilde{x}_{\ell,i},\tilde{y}_{\ell,i})
    \end{aligned}
\end{align}
is chosen, where $[a,b]\approx[c,d]$ means that $|a-c|\leq \frac{1}{2}d_1 $ and $|b-d|\leq \frac{1}{2}d_1 $, and  $\operatorname{val}_{\mathcal{ P},z} (\tilde{x}_{\ell,i},\tilde{y}_{\ell,i})$ represents the value of $z$ for a  $\tilde{x}_{\ell,i},\tilde{y}_{\ell,i}$ coordinated in the point cloud $\mathcal{ P}$.
\end{enumerate}

\subsection{Similarity measure}

The goal is to find $Q_\ell$, $\ell =\{1,\dots, \binom{N}{m+1}\}$ as well as a rotation angle $\theta \in [0,2\pi]$ satisfying requirements \eqref{ineq:req_z} and \eqref{ineq:req_y_theta} which should be similar to $T_\theta$. 
The measure of similarity of the pattern $Q_\ell$ and $T_\theta$ is defined by the Wasserstein distance (see e.g. \cite{santambrogio2015optimal}), which is defined by:
\begin{align*}
    &W_2(U, V) = \min_{\pi}\left( 
    \sum_{i=0}^m \sum_{j=0}^m \pi_{ij} \| u_i - v_j \|_2^2\right)^{1/2},\\
    &U=\{u_i=(u_i^x,u_i^y,u_i^z),\ i=0,\dots,m \}, \quad V=\{v_i=(v_i^x,v_i^y,v_i^z),\ i=0,\dots,m \}
\end{align*}
subject to:
\begin{itemize}
    \item \( \pi_{ij} \geq 0 \),
    \item \( \sum_{j=0}^m \pi_{ij} = \frac{1}{m+1} \) for all \( i = 0, \dots, m \),
    \item \( \sum_{i=0}^m \pi_{ij} = \frac{1}{m+1} \) for all \( j = 0, \dots, m \),
\end{itemize}
where \( \pi = [\pi_{ij}] \) is the transport plan, and \( \| u_i - v_j \|_2 \) denotes the Euclidean distance in \( \mathbb{R}^3 \).

Alternatively,  the Least Squares measure of is considered:
\begin{equation*}
    L(U,V)=\sqrt{\sum_{i=0}^m \|u_i-v_i\|_2^2},
\end{equation*}

\subsubsection{Procrustes distance}

To further assess the influence of rigid transformations on pattern similarity, a Procrustes-type distance is also employed (see \cite{goodall1991procrustes}). Let
\[
\bar{p}^\theta = \frac{1}{m+1}\sum_{i=0}^m p_i^\theta,
\qquad
\bar{q} = \frac{1}{m+1}\sum_{i=0}^m q_i
\]
denote the centroids of $T_\theta$ and $Q$, respectively. We define the centered configurations
\[
\tilde{P}_i^\theta = P_i^\theta - \bar{P}^\theta,
\qquad
\tilde{q}_i = q_i - \bar{q}.
\]

The Procrustes distance is then defined as
\begin{equation}\label{eq:procrustes}
P(T_\theta, Q)
=
\min_{R \in \mathrm{SO}(3)}
\left(
\sum_{i=0}^m
\| \tilde{p}_i^\theta - R \tilde{q}_i \|_2^2
\right)^{1/2},
\end{equation}
where $\mathrm{SO}(3)$ denotes the group of three-dimensional rotation matrices.

In practice, the minimization in \eqref{eq:procrustes} admits a closed-form solution based on the singular value decomposition of the cross-covariance matrix between the centered point sets. Unlike the least-squares distance, the Procrustes distance is invariant with respect to rigid rotations and translations of the candidate pattern.

In the context of this study, the rotation $\theta$ of the reference pattern is already explicitly optimized. Consequently, the Procrustes distance primarily compensates for residual rotational and translational discrepancies induced by discretization and nearest-neighbor projection onto the DEM. Empirically, this measure turns out to be particularly effective in suppressing spurious local minima and yields highly stable estimates of the pattern center.

\subsubsection{Objective function}
Given the reference  pattern \( T \) and a set of target point patterns \( Q_\ell \), $\ell\in \mathcal{ L} \subset  \{1,\dots,\binom{N}{m+1}\}$ satisfying  requirements \eqref{ineq:req_z} and \eqref{ineq:req_y_theta} constructed from the point cloud \( \mathcal{P} \), the goal is to find the point pattern \( Q_{\bar{\ell}} \) and the rotation angle \( \theta \) that minimize the the considered measure between \( T_\theta \) (the rotated version of \( T \)) and \( Q_{{\ell}} \), $\ell\in \mathcal{ L}$. This can be formulated as the following optimization problem:
\begin{equation}\label{opt:W1}\tag{Opt:gen}
    \min_{\theta \in [0, 2\pi]} \min_{\ell \in \mathcal{ L}} \mu(T_\theta, Q_\ell),
\end{equation}

where:\begin{itemize}
\item $\mu$ is considered measure, i.e., Wasserstein $W_2$ or  Least Squares $L2$ or Procrustes $P$.
    \item \( T_\theta \) is the rotated version of \( T \) by angle \( \theta \),
    \item \( Q_\ell \), $\ell \in \mathcal{ L}$ are the point patterns constructed from \( \mathcal{P} \), which in practical situations is the local digital model of terrain (local DEM).
\end{itemize}

In the case of using Least Squares $L2$ measure the optimization problem takes the form
\begin{equation}\label{opt:L2}\tag{Opt:2}
    \min_{\theta \in [0, 2\pi]} \min_{\ell \in \mathcal{ L}} L_2(T_{\theta},Q_{\ell}).
\end{equation}

In the case of using Procrustes measure $P$ the optimization problem takes the form
\begin{equation}\label{opt:P}\tag{Opt:3}
    \min_{\theta \in [0, 2\pi]} \min_{\ell \in \mathcal{ L}} P(T_{\theta},Q_{\ell}).
\end{equation}

To find the  target pattern $Q_{\bar{\ell}}$ the following algorithm is proposed.

\section{Algorithm}\label{sec:algorithm}

Before applying the algorithm, the data requires some preprocessing. Normalization allows data of different ranges to be transformed to a common range, reducing the effect of range differences and controlling data dispersion.

To account for the significant differences in range between spatial coordinates $(X, Y)$ and elevation $(Z)$, all point cloud coordinates are normalized by centering them at the centroid and dividing by their standard deviation. This ensures that each axis contributes comparably to distance-based metrics such as the Wasserstein distance. This normalization prevents our matching algorithm from being dominated by the absolute magnitude of the horizontal coordinates. Thus the normalized representation reflects better the intrinsic geometric structure of the surface patch.

\subsection{Normalization}

Let
$$
\mu = \frac{1}{N} \sum_{i=1}^N (x_i, y_i, z_i)
$$

be a centroid (mass) of point cloud. For each point of the point cloud $\mathcal{ P}$

$$
(x_i', y_i', z_i') = (x_i - \mu_x, y_i - \mu_y, z_i - \mu_z)
$$

the standard deviation is calculated

$$
\sigma_x = \sqrt{\frac{1}{N} \sum_{i=1}^N (x_i')^2}, \quad
\sigma_y = \sqrt{\frac{1}{N} \sum_{i=1}^N (y_i')^2}, \quad
\sigma_z = \sqrt{\frac{1}{N} \sum_{i=1}^N (z_i')^2}
$$

and the normalization of cloud $\mathcal{ P}$ is performed

\begin{equation}\label{normalization}
    (\hat{x}_i, \hat{y}_i, \hat{z}_i) = \left( \frac{x_i'}{\sigma_x}, \frac{y_i'}{\sigma_y}, \frac{z_i'}{\sigma_z} \right)
\end{equation}

\subsection{Scaling of $z$}

The penalty parameter $\lambda>0$  
\begin{equation}\label{normalization_sc}
    (\doublehat{x}_i, \doublehat{y}_i, \doublehat{z}_i) := \left( 
    \hat{x}_i, \hat{y}_i, \lambda\hat{z}_i
    \right)
\end{equation}

has been introduced.

\subsection{Construction Steps / Algorithm}
Given $T= \{ P_i = (x_i, y_i, z_i) \in \mathbb{R}^3 \mid i = 0, \dots, m \}$ and parameter $\lambda>0$. 
The construction of \( Q_\ell \) involves the following steps:

\begin{enumerate}

    \item \textbf{Selection of Points in center:}
    Points \( q_{l,0} := (\tilde{x}_{l,0}, \tilde{y}_{l,0}, \tilde{z}_{l,0}) \) are selected from \( \mathcal{P} \) such that they lie within a radius \( r \) of the center \( q_{l,0} \) (requirement \eqref{ineq:req_z}):
    \[
    \| (x_{0}, y_{0} ) - (\tilde{x}_{l,0}, \tilde{y}_{l,0}) \|_{\infty} \leq r,
    \]
    where \( \| \cdot \|_\infty \) is the $\ell_\infty$ norm,
    and the height constraint satisfy requirement \eqref{ineq:req_z}
    \[
    |\tilde{z}_{l,0} - z_0| \leq 1 \, 
    \]
    where \( z_0 \) is the \( z \)-coordinate of the center of \( T \). These points $q_{l,0}$ serve as centers constructing \( Q_\ell \).
    \item \textbf{Discretization of Angles:}
    The interval \([0, 2\pi]\) is discretized into a finite set of angles \( \theta_1, \theta_2, \dots, \theta_n \).
    \item \textbf{Definition of $Q_{l,\theta}$}:
    For each $q_{l,0}$ satisfying step 1 and for each angle $\theta_i\in \{\theta_1,\dots,\theta_n \}$ we define a target $Q_{l,\theta}$
    as
    \begin{align*}
        Q_{l,\theta}:= \{&  (\tilde{x}_{j}, \tilde{y}_{j}, \tilde{z}_{j})\in \mathcal{ P},\ j=0,\dots,m:\\\
        & (\tilde{x}_{j}, \tilde{y}_{j})= \arg\min_{(x,y) \in \mathcal{P}_{x,y}} \| (x,y)-(x_j',y_j') \|_2,\ 
         \tilde{z}_j= \operatorname{val}_{\mathcal{ P},z} (\tilde{x}_j,\tilde{y}_j) \ j=0,\dots,m,\\
         &\text{where}\  [x_{j}', y_{j}']^T=  [\tilde{x}_{\ell,0},\tilde{y}_{\ell,0}]^T + \operatorname{rot}(\theta) \cdot [x_j-x_0,y_j- y_0]^T 
        \}, 
    \end{align*}
    where $\mathcal{P}_{x,y}$ is the  cloud of $x,y$ coordinates from cloud $\mathcal{ P}$. The above construction of $Q_{l,\theta}$ ensures that $Q_{l,\theta}$ satisfy requirement \eqref{ineq:req_y_theta}.     For given center $q_{\ell,0}$, the points $(x_{j}', y_{j}')$ serves as approximation of rotated reference  pattern $T$ (i.e., $T_\theta$) with respect to center $q_{\ell,0}$, points $(\tilde{x}_j,\tilde{y}_j)$ are the closest points to $(x_{j}', y_{j}')$ in point cloud $\mathcal{ P}$ (with respect to the $x,y$ coordinates) and  $\operatorname{val}_{\mathcal{ P},z} (\tilde{x}_j,\tilde{y}_j)$ represents the value of $z$ for a  $\tilde{x}_j,\tilde{y}_j$ coordinated in point cloud $\mathcal{ P}$.

    \item \textbf{Computation of the considered Similarity Measure ($W_2$, $L_2$ or $P$)}
    For each $Q_{l,\theta}$ from step 3, 
   the considered measure is computed \( \mu(\doublehat{T}_{\theta_i}, \doublehat{Q}_{l,{\theta_i}} )\), where $\doublehat{T}_{\theta_i}$, $\doublehat{Q}_{l,{\theta_i}}$ are obtained from ${T}_{\theta_i}, {Q}_{l,{\theta_i}} $ 
after normalization and scaling according to formulas \eqref{normalization} and  \eqref{normalization_sc}. 

    \item \textbf{Minimization:}
    Find \( Q_{\bar{\ell}} := Q_{\bar{l},\bar{\theta}} \)  which minimizes the considered measure:
    \[
    (\bar{l},\bar{\theta}) =\operatorname{argmin}_{\ell,\theta_i} \mu(\doublehat{T}_{\theta_i}, \doublehat{Q}_{\ell,{\theta_i}}).
    \]

\end{enumerate}

\section{Computational experiments}\label{sec:computional}

The aim of this computational experiment is to test the performance of the considered algorithm together with different measures and different scaling parameters $\lambda$ for finding given target patterns. 
Given the point cloud $\mathcal{ P}$ - DEM numerical 1 m model of terrain in Poland 
in the area $[511511.48,512511.48]\times[216199.39,217199.39]$,  target pattern $T= \{ P_i = (x_i, y_i, z_i) \in \mathbb{R}^3 \mid i = 0, \dots, m \}$ and scaling parameter $\lambda$ - in this computational experiment  $\lambda
\in \{ 1,20,40,60,80,100,200\}$ were chosen. 
\subsection{Checking the operation of the algorithm for matching discrete surface models}

Given the point cloud $\mathcal{ P}$ – digital elevation model of $d_2$ spatial resolution, reference pattern $\mathcal{ P}$ in the form of the crosses defined above  and the parameter $\lambda>0$. In this computational experiment $d_2 = 1$ m reflecting the resolution of DEM for Poland was applied (Section \ref{subsec:high_resolution_DEM}). Moreover, for the purpose of this experiment $\lambda \in \{1,20,40,60,80,100,200\}$ were chosen.

Our test consists of the following steps. We choose the initialization pattern $T_0$, target pattern $T$ and perturbed pattern $\tilde{T}$  by applying the following procedure:
\begin{enumerate}
    \item From the random center point $P_0=(y_0,x_0,z_0)$ taken from an area $[511511.48,512511.48]\times[216199.39,217199.39]$ together with $z_0$ as value of DEM at this cordinate $(y_0,x_0)$, the "cross" is built as the structure of $401$ points, each arm consisting of $100$ points in four sequentially perpendicular directions (anti-clockwise), first direction has an angle of $\alpha$ degrees (random angle in $[0,360)$) to North direction (anti-clockwise) and each direction has 100 points of measurements between 1 meters (up to last entry 100 metre from the center). 
    This structure is referred as an  initialization pattern $T_0$.
     \item For each arm of initialization pattern $T_0$ on the arm:
        \begin{enumerate}
            \item $(y_k,x_k,z_k)$, $k\in \{1,\dots,99\}$,
            \item $(y_k,x_k,z_k)$, $k\in \{101,\dots,199\}$,
            \item $(y_k,x_k,z_k)$, $k\in \{201,\dots,299\}$,
            \item $(y_k,x_k,z_k)$, $k\in \{301,\dots,399\}$
        \end{enumerate}
        To measure the ratio of increase in $z$ to the increase of $(x,y)$ on curve, we compute a discrete arc-length derivative of the height profile using a central finite-difference scheme normalized by the chord length, following  constructions for derivatives along discrete curves,
        \begin{equation*}
            (\Delta_{(x,y)} z)_k=
            \left\{ \begin{array}{ll}
               \frac{z_{k+1}-z_{k-1}}{\sqrt{ (y_{k+1}-y_{k-1})^2+ (x_{k+1}-x_{k-1})^2}},  &  \text{for}\ k\in \{2,\dots,399\} \setminus \{ 100,101,200,201,300,301\}, \\
                \frac{z_{k+1}-z_{0}}{\sqrt{ (y_{k+1}-y_{0})^2+ (x_{k+1}-x_{0})^2}}, &  \text{for}\ k\in \{ 1,101,201,301\}.
            \end{array}
             \right.
        \end{equation*}
        We take $9$ indices ($k$) in each group: $\{1,\dots,99\}$,  $\{101,\dots,199\}$,  $\{201,\dots,299\}$, $\{301,\dots,399\}$  of highest value $(\Delta_{(x,y)} z)_k$ as \textit{exact indices}. The indices $0,100,200,300,400$ we also consider as \textit{exact indices}  (i.e., center and last points of arms). The points $(y_k,x_k,z_k)$, for $k$ in exact indices forms a target pattern $T$\footnote{Note that pattern $T$ consists of 41 points.}.
  
    \item The $yx$ values of the target pattern $T$ obtained in step 2. are translated by random vector $\vec{v}=[v_x,y_y]$ where $v_x,v_y \in [-400,400]$. In this way we get the perurbed pattern $\tilde{T}$.

\end{enumerate}

\subsection{Results}
\begin{figure}[H]
            \includegraphics[width=0.6\linewidth]{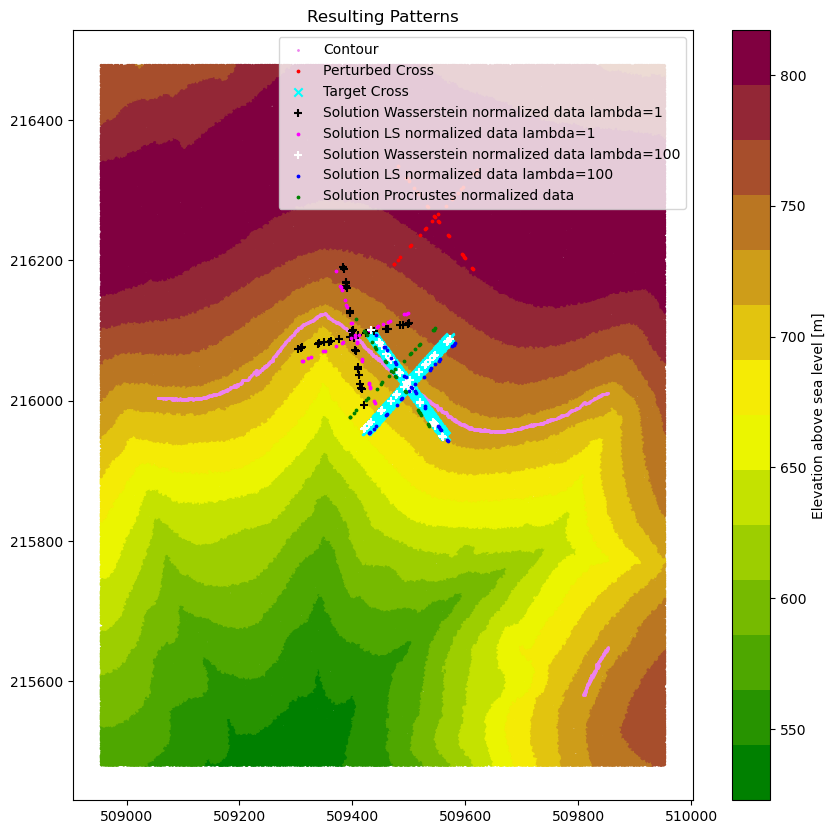}
        \caption{Patterns on map}
        \label{fig:patterns}
\end{figure}

The results of performing $8$ experiments are shown in Table \ref{tab:1} and Table \ref{tab:2} and for the choice of experiment no. 2, the resulting patterns for normalized data are displayed in Figure \ref{fig:patterns} together with the arms profiles in Figure \ref{fig:patterns}.

\begin{table}[H]
\begin{tabular}{rccc}
No. & target cross center & perturbed cross center (Section \ref{sec:computional}) & euclidean distance between \\
1 & (509489.09 , 215942.17) 
 &  (509457.09	, 215833.17) & 113.60

 \\
2 &  (509547.09 , 216261.17) & (509399.09 , 216096.17) & 242.99

\\
3 & (509468.09 , 216009.17) & (509615.09 , 216344.17) & 365.83

\\
4 & (509410.09 , 215979.17) & (509204.09 , 216100.17) & 238.90

\\
5 & (509470.09 , 215992.17) & (509305.09 , 216319.17) & 366.27
\\
6 & (509497.09 , 216005.17) & (509617.09 , 215997.17) & 120.26
\\
7 & (509496.09 , 215956.17) & (509351.09 , 215898.17) & 156.16\\
8 &  (509437.09 , 215939.17) & (509785.09 , 216062.17) & 369.09
\end{tabular}
\caption{The table represents the numerical experiment:\\ \textit{target center} - position of center of the original pattern,\\
\textit{perturbed cross center } - position of ceneter of perturbed  pattern,\\
\textit{euclidean distance between} - the value of euqlidean distance between target cross center and perturbed cross center.}\label{tab:1}
\end{table}
\begin{table}[H]
\begin{tabular}{rrrrrrrrr}
No. & ${W}_1$  & $\tilde{W}_{1}$ & $\tilde{W}_{20}$ & $\tilde{W}_{40}$ & $\tilde{W}_{60}$ & $\tilde{W}_{80}$ & $\tilde{W}_{100}$ & $\tilde{W}_{200}$\\
1 &  38.18& 38.18&  7.81 & 7.81 & 7.81& 7.81& 7.81 & 7.81  \\
2 & 123.00 &  118.22 &  33.83  & 1.41& 1.41  &  1.41& 1.41 &  1.41
\\
3 & 121.03 &  106.04 &  4.47  & 4.47 & 4.47 & 4.47& 4.47 & 4.47
\\
4 & 89.89 &  106.92 &  122.03  & 11.31&  11.31& 11.31 & 11.31& 11.31\\
5 & 155.97 &  161.83 &  63.41  & 229.10& 10 & 10
 & 10 & 25.23 \\
6 & 131.52 &  121.31 &   2 & 2&  2 & 2 & 2 & 2
  \\
7 & 70.21 &  81.60 &  3.60  & 3.60& 3.60& 3.60& 3.60 & 3.60\\
8 & 371.38 &  371.38 &  317.80  & 1& 1 & 1 & 1 & 1
\end{tabular}\\[0.5cm]
\begin{tabular}{rrrrrrrrr}
No. & $\tilde{LS}_1$ & $\tilde{LS}_{20}$ & $\tilde{LS}_{40}$ & $\tilde{LS}_{60}$& $\tilde{LS}_{80}$ & $\tilde{LS}_{100}$ & $\tilde{LS}_{200}$& PC \\
1 &  38.18 & 1.41 & 1.41& 1.41 & 1.41& 1.41 & 1.41&1.41 \\
2 & 115.20 & 17.20 & 17.20& 17.20 & 17.20 & 9.43 &9.43& 31.90
\\
3 & 106.04 & 4.47 & 4.47& 4.47 &  4.47& 4.47&4.47& 4.47
\\
4 & 106.92 & 12.04 & 1.41& 1.41 & 1.41  & 1.41&1.41& 1\\
5 & 161.83 & 27.01 &10  & 10
& 10
 &  10 & 10& 7.07\\
6&  121.31  & 2.82 & 2.82& 2.82 &  2.82& 2.82 & 2.82 &2.82\\
7 & 81.60 &  18.02 & 22.47 & 22.47&  22.47&  22.47 & 425.77& 419.31\\
8 & 371.38 & 1 & 1& 1 &  1
 & 1 & 1&1
\end{tabular}
\caption{The table represents the numerical experiment:\\ 
$W_1$ - non-normalized data, euclidean distance between center of original pattern and solution pattern obtained with algorithm based on Wasserstein distance with $\lambda=1$\\
$\tilde{W}_{\lambda}$ - normalized data, euclidean distance between center of original pattern and solution pattern obtained with algorithm based on Wasserstein measure with choosen $\lambda$\\
$\tilde{LS}_{\lambda}$ - normalized data, euclidean distance between center of original pattern and solution pattern obtained with algorithm based on Least Squares measure with choosen $\lambda$ \\
$PC$ - normalized data, euclidean distance between center of original pattern and solution pattern obtained with algorithm based on Procrusters measure.
}\label{tab:2}
\end{table}

\begin{figure}[H]
     \centering
     \begin{subfigure}[b]{0.65\textwidth}
         \centering
            \includegraphics[width=\linewidth]{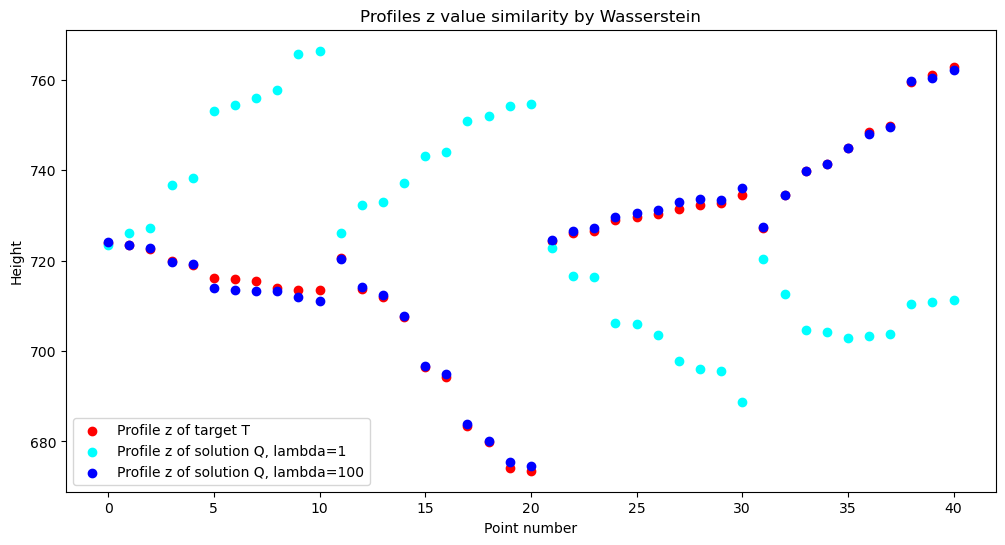}
         \caption{Height profiles of solutions with similarity by Wasserstein distance}
     \end{subfigure}\\
     \begin{subfigure}[b]{0.65\textwidth}
         \centering
         \includegraphics[width=\linewidth]{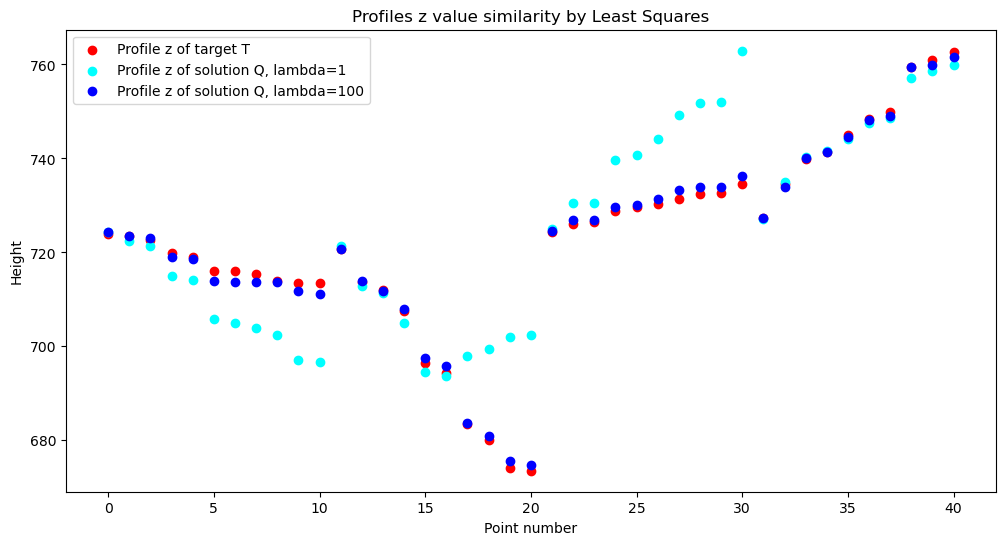}
         \caption{Height profiles of solutions with similarity by Least Squares distance}
     \end{subfigure}\\
      \begin{subfigure}[b]{0.65\textwidth}
     \centering
     \includegraphics[width=\linewidth]{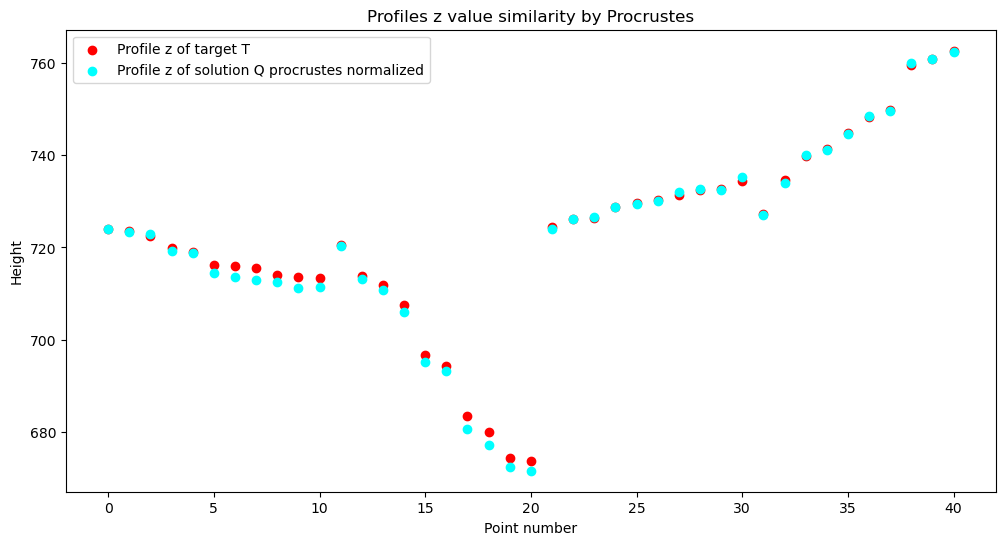}
     \caption{Height profiles of solutions with similarity by Procrustes distance}
 \end{subfigure}
        \caption{Height profiles}
        \label{fig:profiles}
\end{figure}

\section{Acknowledgments}
The article is the result of cooperation between scientific teams from the Systems Research Institute of the Polish Academy of Sciences, Cardinal Stefan Wyszynski University, and the Institute of Geodesy and Cartography. The authors would like to acknowledge PBG Geophysical Exploration Ltd for sharing archival documentation concerning the local topography around selected gravimetric points and their coordinates as well as the Polish Geological Institute - National Research Institute (PGI-NRI) for providing additional information about the local topographic data used and for manual selecting new, more probable locations of test gravimetric points.

\bibliographystyle{plain}
\bibliography{bibliography}

\end{document}